 \documentclass[authoryear,preprint,review,12pt]{elsarticle}

\usepackage{xr}
\usepackage{color}

\usepackage{graphicx}
\usepackage{lineno}  
\usepackage{amssymb}
\usepackage{float}
\usepackage{arydshln}
\usepackage{url}

\usepackage{multirow}
\usepackage{threeparttable}
\usepackage{bm}
\usepackage{amsmath}

\usepackage[x11names]{xcolor}
\usepackage[
  colorlinks,
]{hyperref}

\AtBeginDocument{\hypersetup{citecolor=DodgerBlue4}}

\journal{PNAS}

\begin{document}

\begin{frontmatter}
\title{Pervasive impact of spatial dependence on predictability}

\author[a,b]{Peng Luo}
\ead{pengluo@mit.edu}

\author[c]{Yongze Song\corref{cor1}}
\cortext[cor1]{Corresponding author: Yongze Song, yongze.song@curtin.edu.au}

\author[d]{Wenwen Li}
\ead{wenwen@asu.edu}

\author[b]{Liqiu Meng}
\ead{liqiu.meng@tum.de}

\address[a]{Senseable City Lab, Massachusetts Institute of Technology, Cambridge, USA}

\address[b]{Chair of Cartography and Visual Analytics, Technical University of Munich, Munich, Germany}

\address[c]{School of Design and the Built Environment, Curtin University, Perth, Australia}
\address[d]{School of Geographical Sciences and Urban Planning, Arizona State University, USA}

\begin{abstract}

Understanding the complex nature of spatial information is crucial for problem solving in social and environmental sciences. This study investigates how the underlying patterns of spatial data can significantly influence the outcomes of spatial predictions. Recognizing unique characteristics of spatial data, such as spatial dependence and spatial heterogeneity,  we  delve into the fundamental differences and similarities between spatial and non-geospatial prediction models. Through the analysis of six different datasets of environment and socio-economic variables, comparing geospatial models with non-geospatial models, our research highlights the pervasive nature of spatial dependence beyond geographical boundaries. This innovative approach not only recognizes spatial dependence in geographic spaces defined by latitude and longitude but also identifies its presence in non-geographic, attribute-based dimensions. Our findings reveal the pervasive influence of spatial dependence on prediction outcomes across various domains, and spatial dependence significantly influences prediction performance across all spaces. Our findings suggest that the strongest spatial dependence is typically found in geographic space for environment variables, a trend that does not uniformly apply to socio-economic variables. This investigation not only advances the theoretical framework for spatial data analysis, but also proposes new methodologies for accurately capturing and expressing spatial dependence under complex conditions. Our research extends spatial analysis to non-geographic dimensions such as social networks and gene expression patterns, emphasizing the role of spatial dependence in improving prediction accuracy, thereby supporting interdisciplinary applications across fields such as geographic information science, environmental science, economics, sociology, and bioinformatics.

\end{abstract}

\end{frontmatter}


\section{Introduction}

Spatial data and prediction play a crucial role in modern society, covering key areas such as environmental monitoring, urban planning, and public health \citep{cressie1988spatial,goodchild1993framework}. The rapid development of artificial intelligence (AI) has enabled the reasonable representation and embedding of spatial data and significantly enhanced the performance of spatial prediction tasks \citep{mai2020multi}. Despite these advance, a comprehensive understanding of spatial data, as the key of spatial modeling \citep{goodchild1992geographical}, remains limited. As a unique form of data, spatial data fundamentally differs from other types of data, characterized by two main spatial effects: spatial dependence \citep{anselin1988spatial,miller2004tobler} and spatial heterogeneity \citep{fotheringham2003geographically,goodchild2004validity}. Spatial dependence refers to the correlation between the values at a location and its surrounding locations in geographic space \citep{anselin1995local,luo2022generalized}. On the other hand, spatial heterogeneity describes the variability or diversity of spatial data across different geographic locations, indicating that the distribution, intensity, or pattern of spatial processes or phenomena changes with geographic location, reflecting the non-uniformity of spatial data \citep{getis1992analysis,de2005dealing,fotheringham2003geographically}. Spatial analysis serves as a bridge connecting different disciplines, promoting interdisciplinary research development. Understanding the nature of spatial effects and their impact on the performance of prediction models is crucial for improving the accuracy of spatial prediction tasks \citep{oliver1990kriging}.

Spatial heterogeneity is more readily accepted across various scientific disciplines. Various predictive and explanatory models have been developed for applying spatial heterogeneity, including recent advances in deep learning and machine learning models, which aimed at capturing and fitting the complex relationships within data \citep{goodchild2021replication}. Therefore, spatial heterogeneity is frequently considered in the analysis of spatial data, whether explicitly or implicitly. Conversely, the concept of spatial dependence is less frequently discussed in other scientific fields, making it a more fundamental characteristic that distinguishes spatial data from other data types. First, spatial dependence results in sample points not being independent of each other but statistically correlated. This correlation can reduce the number of effective sample points, making models prone to underfitting \citep{griffith2003spatial}. Moreover, it presents challenges in model validation. Randomly dividing training and test sets can lead to an interspersed distribution of test and training samples in space, overestimating prediction accuracy due to the presence of spatial dependence.

Geographic variables can be presented as data at both geographic space defined by latitude and longitude, and a attribute space composed of other geographic characteristics. We argue that geographic space differs from attribute spaces, primarily due to its inherent property of spatial dependence.
Mathematically, spatial dependence describes the autocorrelation of a variable Z in a space composed of (X, Y). Yet, if (X, Y) are not geographic coordinates but represent any two non-geographic, are the dependencies in spatial dimensions consistent with the attribute dimensions, or are there unique characteristics? In other words, the closer (X, Y) are, the closer the value of Z. This seems to also apply to non-geospatial predictive methods based on attributes. If there is no fundamental difference between spatial and attribute dimension dependencies, the scientific contribution of spatial prediction methods needs to be reevaluated. This question reveals the fundamental importance of spatial data analysis in scientific research and practical applications. Realizing that spatial dependence exists not only in geographic space but can also affect other areas such as social networks, gene expression patterns, and the spatial distribution of economic activities, greatly expands our application range and the depth of spatial data analysis methods.

This study explores the universality and essential characteristics of spatial dependence for a better understanding of the general laws of spatial data. It aims to provide new perspectives to geographic information science and environmental science, and new analytical tools for economics, sociology, and bioinformatics, among others \citep{luo2022generalized}. Being driven by the characteristics of spatial dependence, we attempt to develop novel models and design specific techniques to incorporate this aspect effectively. By determining the optimal geographic or attribute space for spatial prediction, the accuracy of spatial predictions can be improved. This will not only advance the theory of spatial data analysis but also provide more powerful and flexible tools for solving practical problems, especially in scenarios where accuracy and reliability are crucial.

We introduced the concept of pan-spatial statistics, assuming that spatial dependence exists in both geographic space composed of latitude and longitude and non-geographic space composed of any two attributes. We chose two indicators of spatial dependence: the Moran's Index \citep{moran1950notes} and the q statistic \citep{wang2010geographical,luo2021spatial}. The former is a indicator for continuous spatial dependence, based on the correlation of numerical values for each observations with its surrounding observations. The latter was initially proposed to describe the apparent stratified heterogeneity in the spatial distribution of variables, as clear stratification also represents strong dependence within layers, making it considered for assessing stratified dependence \citep{luo2022identifying}. We employed two geospatial models, Ordinary Kriging (OK) and Regression kriging (RK), along with three non-geospatial models, Linear Regression Model (LM), Decision Tree (DT), and Random Forest (RF). They are all widely applied and have achieved good results. Firstly, we performed spatial predictions using attributes of each spatial dimension, then evaluated the accuracy of predictions and the correlation with spatial dependence in that space. Secondly, we analyzed the performance differences in spatial predictions within spaces with the highest spatial dependence, comparing those in geographic space.

The developed methods are implemented in the spatial predictions for six sets of geographic datasets,  comprising three sets of environmental variables and three sets  of socio-economic variables. The environment variable data included the distribution of three trace elements (i.e. Cu, Zn, and Pb), while the socio-economic variable data involved the homeless rates in three major cities in Australia (i.e., Melbourne, Sdyney, and Brisbane). For heavy metal content and homeless rate, we selected 9 and 10 explanatory variables for model training and prediction, respectively. Our results indicate that: 1) Spatial dependence exists not only in geographic space but also in attribute space; 2) Regardless of whether in geographical or non-geographic space, the magnitude of spatial dependence is strongly correlated with the performance of spatial predictions; 3) For environment variables, the largest spatial dependence is likely to exist in geographic space, a rule that does not necessarily apply to socio-economic variables. This may represent one of the most fundamental characteristics of geographic data.

\section{Implementations and Results}

\subsection{Prediction performance and spatial effect}

The concept of pan-spatial statistic is implemented in analyzing the relationship between spatial dependence and spatial prediction accuracy for six datasets, comprising three sets of environment variables and three sets of socio-economic variables. The environmental data consist of distributions of three trace elements (Cu, Pb, and Zn) in a region of Australia. The socio-economic data represent the distribution of homelessness rates in three Australian cities: Melbourne, Sydney, and Brisbane. Detailed data distributions can be found in Supplementary Figure S1. 

In this section, we separately examined the impact of spatial dependence on geospatial models (section 2.1.1) and non-geospatial models (section 2.1.2). For the former, we utilized OK model, while for the latter, we employed LM, DT, and RF.

\subsubsection{Spatial effect in geospatial models}

\begin{figure}[ht!]
\centering\includegraphics[width=1.0\linewidth]{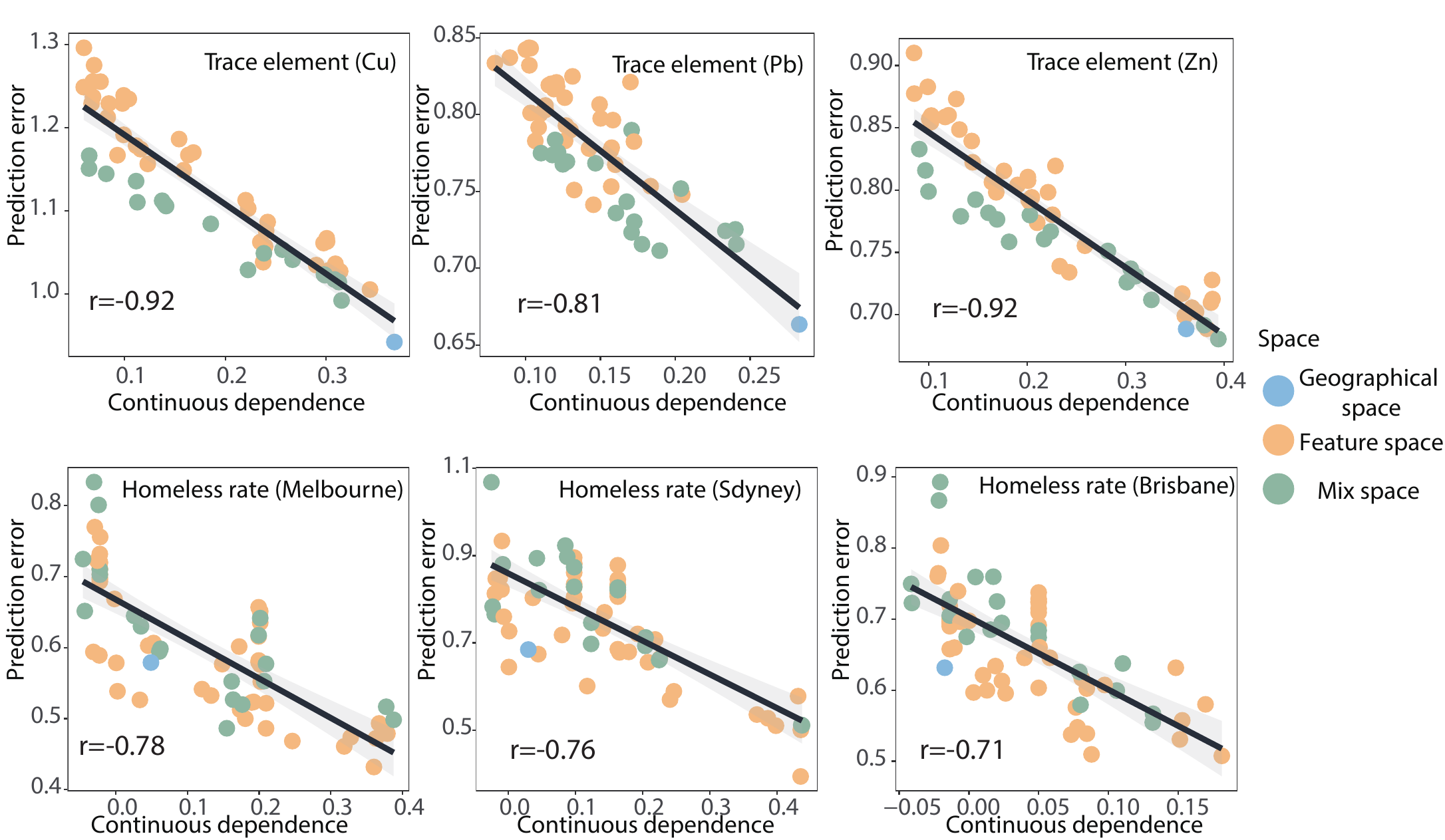}
\caption{The relationship between continuous dependence in geographic and attribute spaces and the predictive performance of the OK model is represented by different colored points. Blue points signify geographic space denoted by longitude and latitude, indicating pure geographical dimensions. Green points represent a mix of spatial dimensions, where one dimension is a geographic space dimension (either longitude or latitude) and the other is a attribute, highlighting the combination of geographical and attribute-based analysis. Orange points denote instances where both spatial dimensions are attributes, illustrating a shift towards entirely attribute-based spatial analysis. This categorization underscores the varying impact of spatial dependence, depending on whether the analysis leans more towards traditional geographic dimensions or incorporates additional attribute dimensions, on the predictive accuracy of the OK model.}
\label{fig:01}
\end{figure}

Figure \ref{fig:01} presents the scatter plot between spatial dependence, measured by Moran's I index and prediction errors of Kriging model across different spatial contexts, including geographic and attribute spaces. The strong negative Pearson correlation between spatial dependence and prediction errors for environmental variables such as trace elements (Cu, Pb, and Zn) suggests an underlying natural order that physical and chemical processes exhibit consistency and predictability within geographic space. In contrast, socio-economic variables, exemplified by homelessness rates in Melbourne, Sydney, and Brisbane, also demonstrate a negative correlation with spatial dependence, albeit to a lesser extent than environment variables. The correlation coefficients in these cases, while still indicative of an inverse relationship, reflect the greater complexity and uncertainty inherent in socio-economic phenomena, which are substantially shaped by human actions, policy decisions, and a myriad of other socio-cultural factors. Consequently, socio-economic variables pose a greater challenge for prediction and explanation through spatial analysis, as they are less bound by the natural laws that govern the distribution of physical and chemical attributes in geographic space.


Figure \ref{fig:01} provides a clear depiction of the differences and connections between geographic and non-geographic spaces in terms of spatial dependence and prediction errors. Environmental variables (e.g. Cu, Pb, and Zn) in geographical space (blue dots in Figure \ref{fig:01}) demonstrate the strongest spatial dependence among all type of space. This observation suggests a high level of order and consistency in how environmental variables are distributed in geographic space, with location playing a key role in predicting these attributes accurately. High spatial dependence for environmental variables within geographic space likely stems from consistent natural processes and physical laws, such as geological composition, soil types, and climatic conditions. These elements often have clear spatial patterns that are not random but instead can be systematically understood and predicted using geographic geospatial models. On the other hand, the attribute space represented by orange dots and the mixed space indicated by green dots show lower spatial dependence for environmental variables. This observation could be because attribute space doesn't directly consider geographic location but is based on other attributes, such as chemical properties or economic indicators, which might not have strong spatial relationships.

For socio-economic variables such as the homelessness rates in Melbourne, Sydney, and Brisbane, the plots also show some level of spatial dependence, but it's not as strong as the environmental variables in geographic space. This could reflect the complex impact of human activities, policy decisions, and social structures on socio-economic variables, which might not display as clear spatial patterns in geographic space.


\begin{figure}[ht!]
\centering\includegraphics[width=1.0\linewidth]{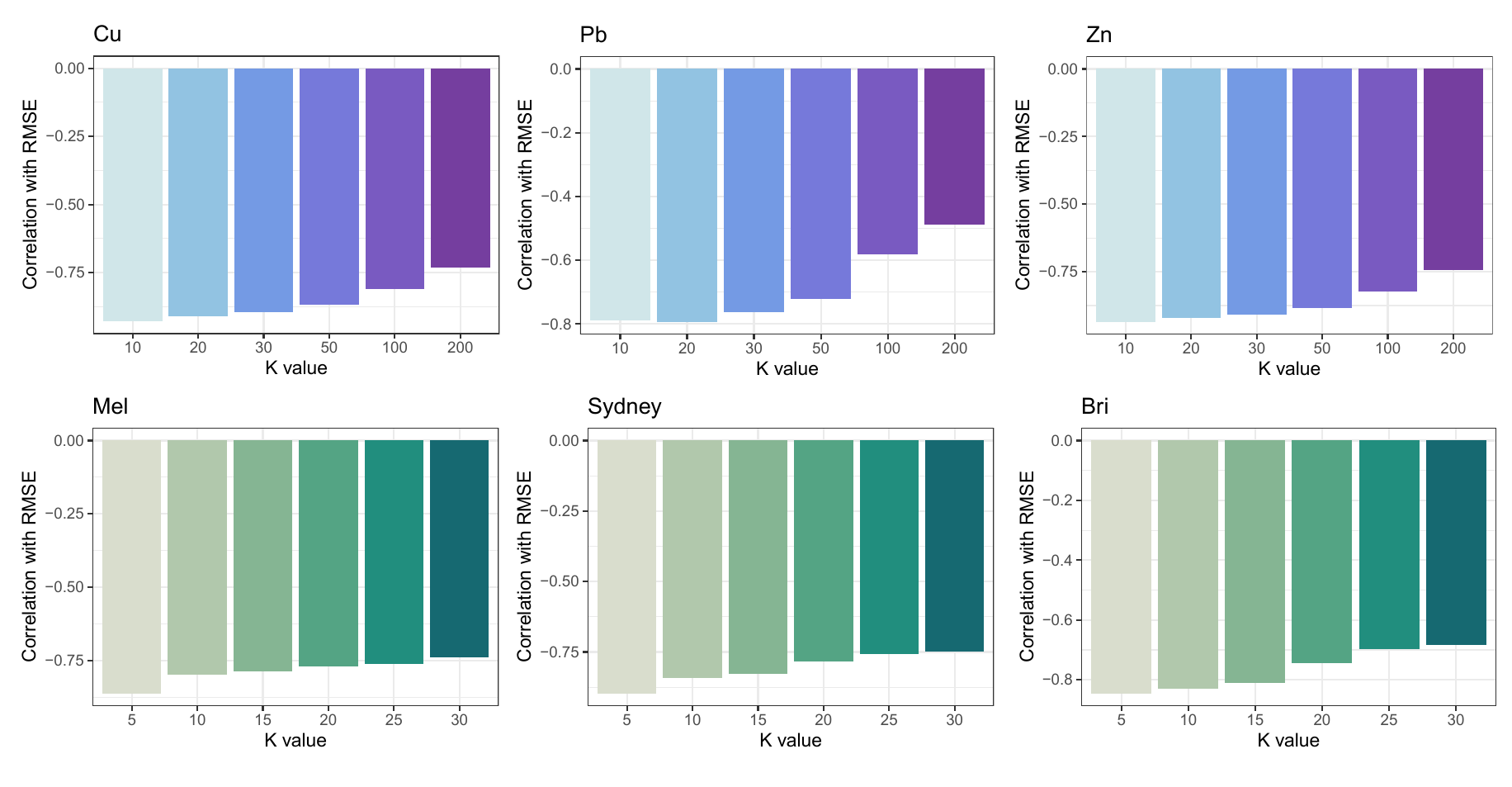}
\caption{The sensitivity analysis of the number of neighboring points, \(K\), for calculating the Moran's Index of local spatial dependence presents findings across various settings of \(K\) (10, 20, 30, 50, 100, 200). We computed the Moran's Index for the entire spatial domain under each \(K\) configuration. Subsequently, the Pearson correlation coefficient between the Moran's Index for all points and the prediction errors of the Ordinary Kriging (OK) model was calculated. This analysis aims to elucidate how the choice of \(K\) influences the ability of the Moran's Index to reflect local spatial dependencies in relation to the accuracy of spatial predictions made by the OK model.}
\label{fig:02}
\end{figure}

We conducted a sensitivity analysis on Moran's I (Figure \ref{fig:02}). The performance of Moran's I is easily influenced by the method of constructing spatial weights. An important parameter in constructing spatial weights is the number of nearest neighbors, K, involved in the calculation. Therefore, we selected different K values to analyze the impact of changes in K on our conclusions. For cases involving environmental variables, with approximately 1000 observations, we chose K values of 10, 20, 30, 50, 100, and 200. For cases involving socio-economic variables, with less than 100 observations, we chose K values of 5, 10, 15, 20, 25, and 30. The results, as shown in Figure \ref{fig:02}, indicate that our conclusions remain stable under different K values: there is a strong negative correlation between the prediction error of the geostatistical model and the magnitude of Moran's index. This correlation can reach above 0.9. For example, when K is 10, the correlation between Moran's index for Cu and Zn distribution and geostatistical error is -0.92 and -0.93, respectively. Additionally, we found that this relationship decreases as K increases. When K increases from 50 to 200, there is a significant drop in the correlation for environmental variables.

\begin{figure}[ht!]
\centering\includegraphics[width=1.0\linewidth]{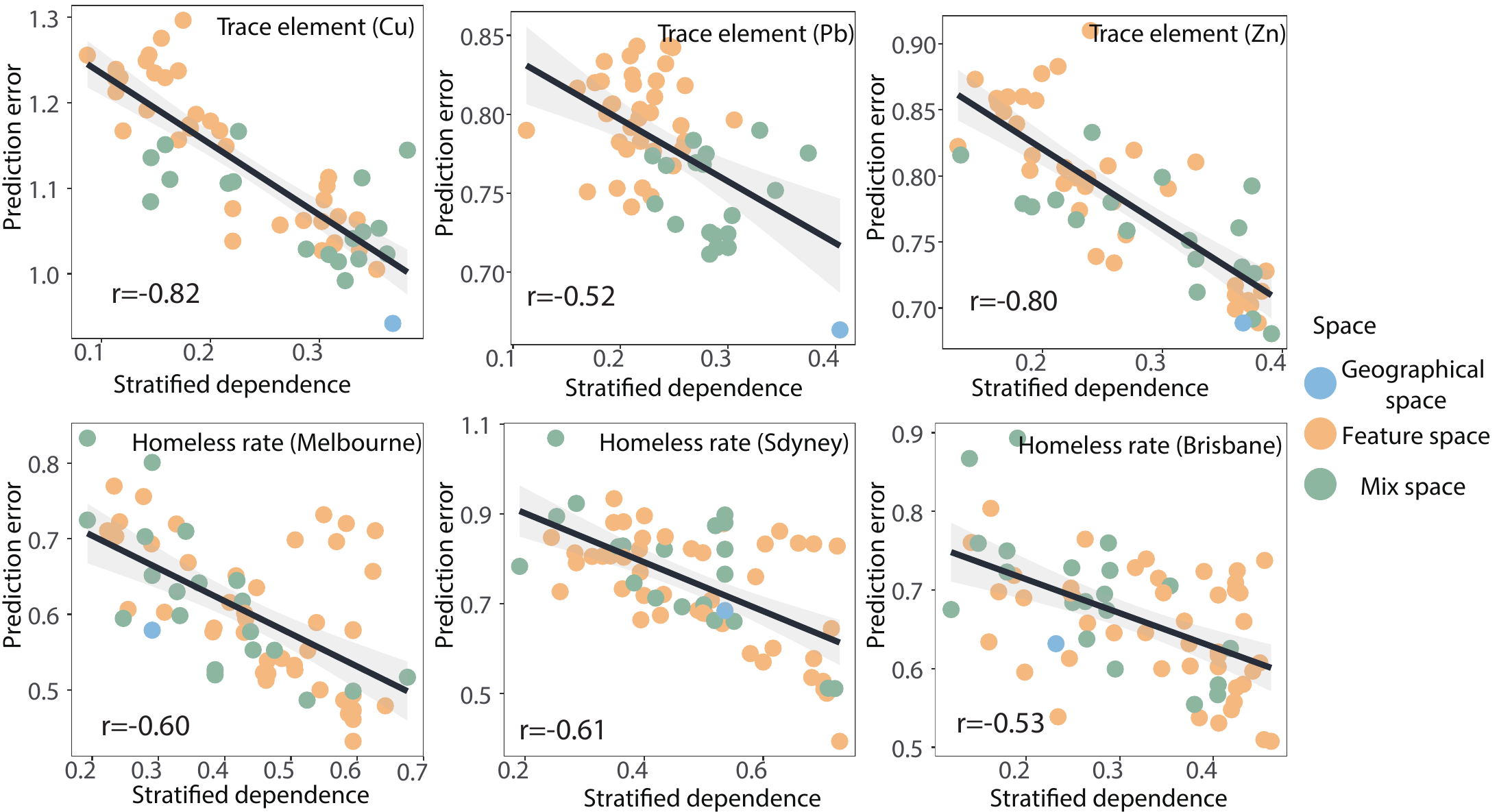}
\caption{The relationship between stratified dependence in geographic and attribute spaces and the predictive performance of the OK model is represented by different colored points. Blue points signify geographic space denoted by longitude and latitude, indicating pure geographical dimensions. Green points represent a mix of spatial dimensions, where one dimension is a geographic space dimension (either longitude or latitude) and the other is a attribute, highlighting the combination of geographical and attribute-based analysis. Orange points denote instances where both spatial dimensions are attributes, illustrating a shift towards entirely attribute-based spatial analysis.}
\label{fig:03}
\end{figure}

Figure \ref{fig:03} displays the relationship between stratified dependence and the prediction errors of the Kriging model in different spaces, including geographic and attribute spaces. SSH can measure spatial dependence on a larger scale (i.e., strata). Across all six cases, there is a strong correlation between the two, although this correlation is generally weaker than the predictive performance correlation with Moran's I index. This may indicate that spatial dependence at a smaller scale is more determinative of spatial prediction performance. Similarly, to the results with Moran's I index (Figure \ref{fig:01}), the correlation between SSH and predictive performance is stronger for environment variables than for socio-economic variables. Finally, the results also reveal that for environment variables, there is a strong stratified dependence in geographic space. In contrast, for socio-economic variables, the stratified dependence in geographic space is lower, with many instances showing stronger dependence in attribute space.

In summary, our results quantitatively demonstrate the impact of spatial dependence on the performance of geostatistical spatial prediction. Additionally, we have found that: first, the control of dependence on predictive performance is not only present in geographic space but exists in any attribute space; second, the impact of dependence on performance is more significant for environment variables. This may be because socio-economic variables are more susceptible to human influence and are more challenging to explain and predict from a spatial perspective compared to environment variables; third, our findings provide evidence for the continued popularity of geostatistical methods in the spatial prediction of environment variables: for the spatial distribution of environment variables, their numerical values have the strongest dependence in geographic space represented by latitude and longitude.

\subsubsection{Spatial effect in non-geospatial models}

After demonstrating the relationship between spatial dependence and the prediction performance of geo-model (i.e. kriging), we further explore the whether this conclusion exists in non-geospatial prediction models. We selected three non-geospatial models to perform spatial predictions in two cases, namely the natural geographic variable Pb and Sydney's homeless risk. In this experiment, these three models conduct the prediction based on two variables that calculate spatial dependence. During training, we set 80\% as the training set ratio, repeated 50 times, and took the average RMSE.

\begin{figure}[ht!]
\centering\includegraphics[width=1.0\linewidth]{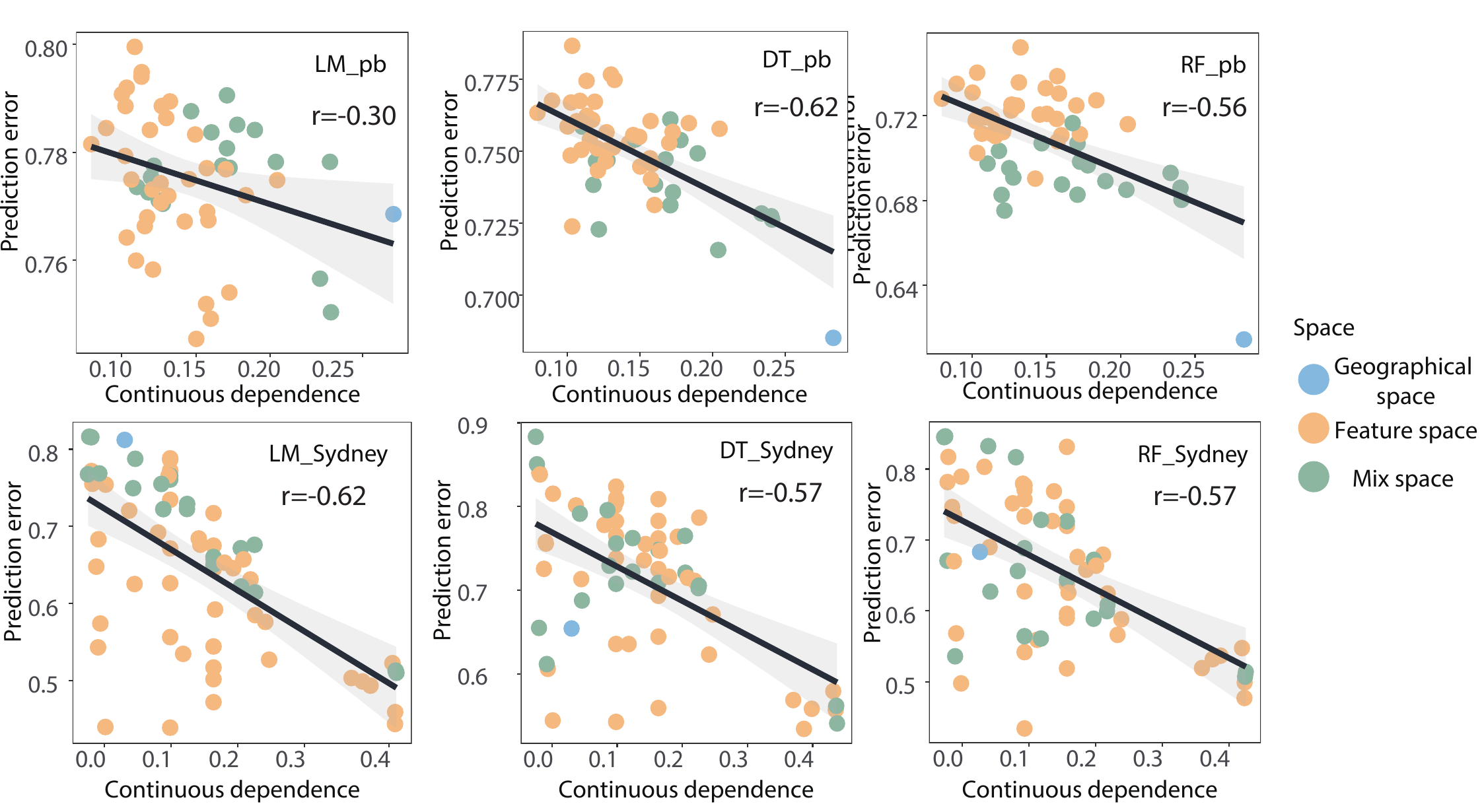}
\caption{The relationship between continuous dependence in geographic and attribute spaces and the predictive performance of three non-geospatial models is examined. The models in question are LM, DT, and RF. In predictions, these models solely utilize two spatial dimensions consisting of geographic variables (longitude and latitude) and attributes. Blue points denote geographic space represented by longitude and latitude, emphasizing the use of purely geographical dimensions for modeling. Green points signify a combination of spatial dimensions, where one is a geographic space dimension (either longitude or latitude) and the other is a attribute, representing an integration of geographic and attribute-based information. Orange points indicate that both spatial dimensions are attributes, illustrating a fully attribute-based approach in spatial analysis. This configuration highlights the varying influences of continuous dependence on the predictive accuracy of the LM, DT, and RF models, depending on the composition of geographic and attribute spaces utilized.}
\label{fig:04}
\end{figure}

\begin{figure}[ht!]
\centering\includegraphics[width=1.0\linewidth]{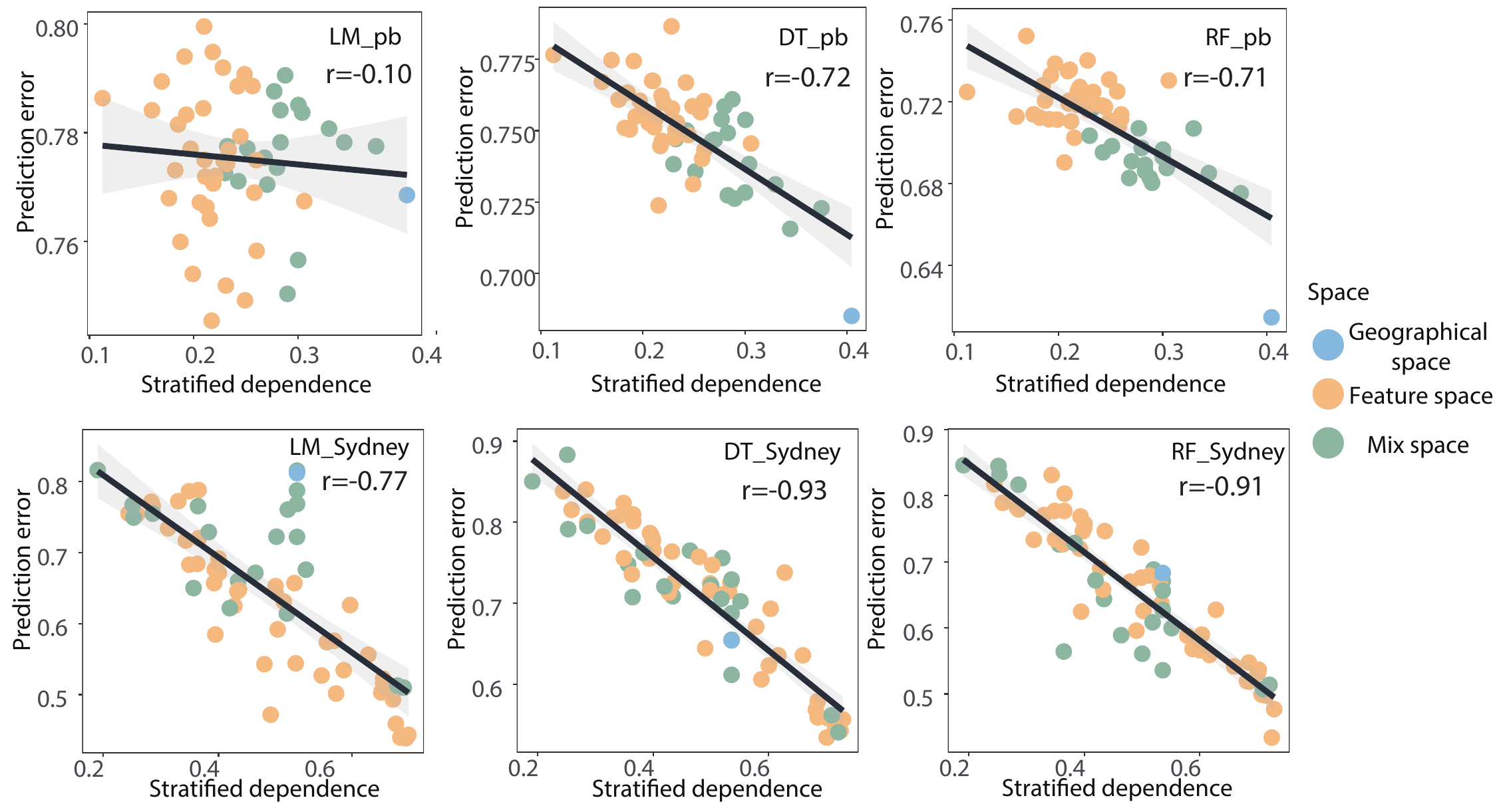}
\caption{The relationship between stratified spatial dependence in geographic and attribute spaces and the predictive performance of three non-geospatial models is examined. The models in question are LM, DT, and RF. In predictions, these models solely utilize two spatial dimensions consisting of geographic variables (longitude and latitude) and attributes. Blue points denote geographic space represented by longitude and latitude, emphasizing the use of purely geographical dimensions for modeling. Green points signify a combination of spatial dimensions, where one is a geographic space dimension (either longitude or latitude) and the other is a attribute, representing an integration of geographic and attribute-based information. Orange points indicate that both spatial dimensions are attributes, illustrating a fully attribute-based approach in spatial analysis.}
\label{fig:05}
\end{figure}

Figure \ref{fig:04} and Figure \ref{fig:05} present the results for the three non-geospatial models for predicting two geographical variables. The results show that although not as strong as geostatistical models, spatial effects can also significantly affect the performance of spatial predictions. For the continuous dependence (Figure \ref{fig:04}), the correlation between dependence and prediction accuracy for the decision trees and random forests on Pb were -0.62 and -0.56, respectively, and on homeless risk were both -0.57. The LM model's predictive performance had a weaker relationship for Pb, with correlation on Pb and homeless risk being -0.30 and -0.62, respectively.

The relationship between predictive performance and stratified dependence (i.e. q statistic) is stronger, especially for tree-based models (Figure \ref{fig:05}). For decision tree and random forest models, the correlation between stratified dependence and predictive performance ranges between 0.71 and 0.93, significantly higher than the results for the Kriging model. Additionally, their prediction accuracy also surpasses that of the Kriging model. This may reveal that tree-based models are more adept at capturing stratified dependence, thereby achieving higher accuracy. In contrast, geo-models, such as Kriging, tend to capture local spatial dependencies. This is because their predictive performance relies on the construction of the semivariogram, which is built based on pairs of sample points, hence more indicative of continuous dependence. For the LM model, its performance correlation with stratified dependence is not significant in the Pb case (r=-0.10), but stronger in the homeless rate case (r=-0.77). This outcome is consistent with the conclusions on continuous dependence (with correlation coefficients for Pb and homeless rate respectively at -0.30 and -0.62).

\subsection{Spatial dependence controls the accuracy of spatial prediction }

\subsubsection{Performance in geospatial models}

In this section, we explore how spatial dependence can quantitatively improve the performance of spatial predictions. Having demonstrated a remarkable correlation between spatial dependence and the performance of spatial predictions, we can improve the accuracy of spatial prediction tasks by judiciously leveraging this finding. Specifically, by assigning greater weight to the spaces with the strongest spatial dependence, whether in geographic space or attribute space, we hope to more fully utilize the characteristics of geographic data, thereby enhancing the accuracy of spatial predictions.

\begin{figure}[ht!]
\centering\includegraphics[width=1.0\linewidth]{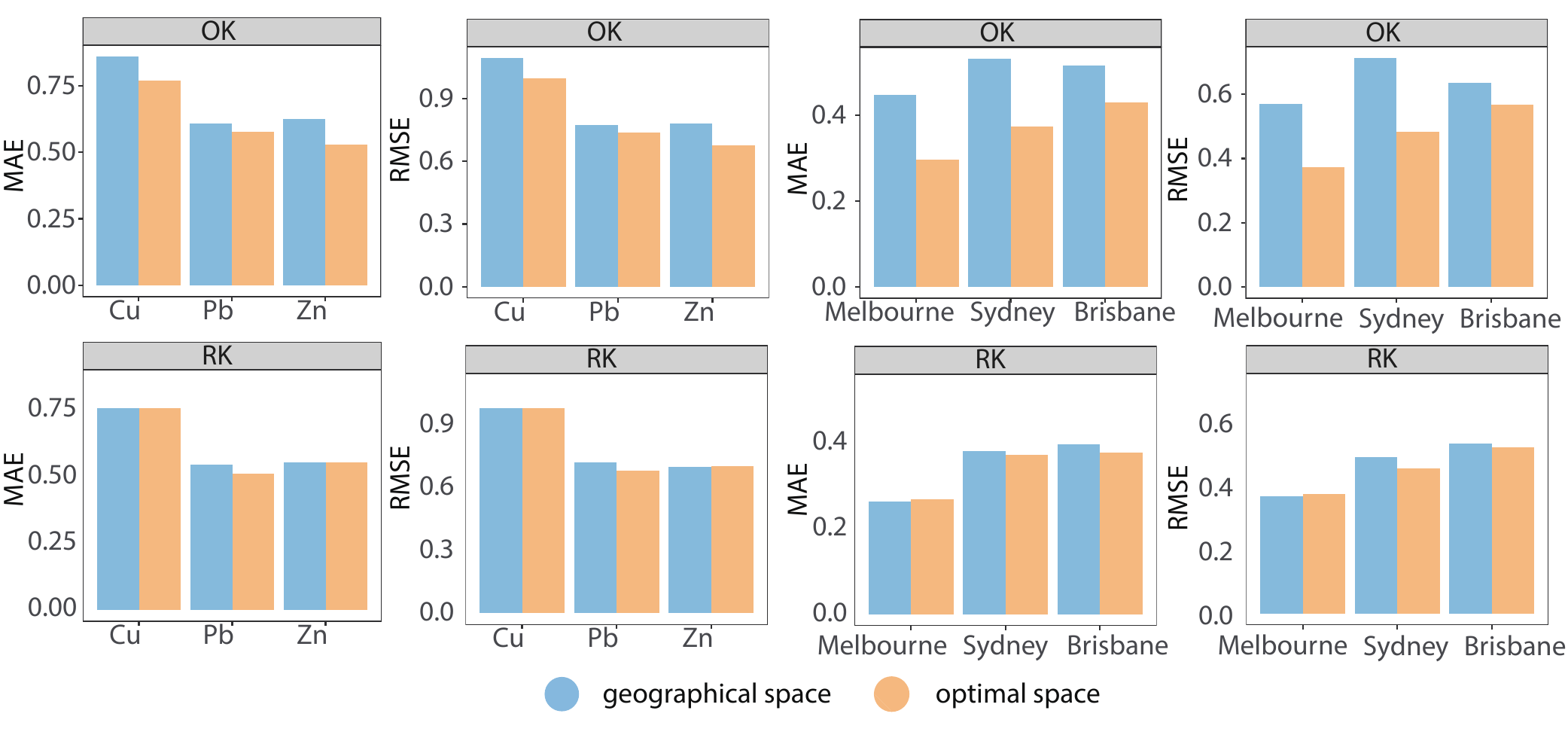} 
\caption{For geospatial models, predictive performance is compared between geographical space and optimal space (the space with the strongest spatial dependence). The blue bars represent the results in the geographical space, while the orange bars represent those in the optimal space. For OK, only two dimensions of features selected in the space are used for interpolation. In contrast, for RK, interpolation is performed using all features after trend surface fitting, followed by interpolation using the selected spatial dimensions.}
\label{fig:06}
\end{figure}

For the two geospatial models, Ordinary Kriging and Regression Kriging, the key to performing spatial predictions lies in constructing the semivariogram of geographic data in space. Since we have demonstrated that spatial dependence exists not only in geographic space but also in attribute space, we have chosen the attribute space with the strongest spatial dependence, specifically the one with the highest Moran's I index, for conducting spatial predictions. Simultaneously, we also ran both models in geographic space to compare the degree of performance improvement in the optimal space versus geographic space. Figure \ref{fig:06} and Table \ref{table1} shows the results of the accuracy comparison, indicating that in most cases, spatial predictions performed in the optimal space see an improvement in performance. For OK, in all six datasets, the prediction accuracy in the optimal space is higher than in geographic space, with the Mean Absolute Error (MAE) being 5.633\% to 50.444\% higher. Moreover, the improvement is more pronounced for socio-economic variables. For the prediction accuracy of the homeless rate in Melbourne, Sydney, and Brisbane, there is an increase of 50.444\%, 42.574\%, and 20.176\% respectively. This is because, as shown in Figure \ref{fig:01}, the spatial dependence in the optimal attribute space is significantly higher for socio-economic variables than in geographic space. In the case of environment variables, the attribute space with the highest spatial dependence is either slightly inferior to or close to geographic space, hence the prediction performance in the optimal attribute space and geographic space is similar. For RK, the performance improvement in the optimal attribute space is marginal, and even shows a slight decrease in the cases of Zn and Melbourne. This might be because RK utilizes all attributes to fit the trend, then predicts residuals over the area with OK. In such instances, the difference in spatial dependence between the optimal space and geographic space is somewhat mitigated.

\begin{table}[!ht]
    \centering
    \caption{Performance increment}
    \scriptsize
\begin{tabular}{cccl|cr}
\hline & & \multicolumn{2}{c|}{ Ordinary Kriging } & \multicolumn{2}{c}{ Regression Kriging } \\
\hline \multicolumn{2}{c}{ Variable } & \multicolumn{1}{c}{ MAE $\downarrow$} & \multicolumn{1}{c|}{ RMSE $\downarrow$} & \multicolumn{1}{c}{ MAE $\downarrow$} & \multicolumn{1}{c}{ RMSE $\downarrow$} \\
\hline \multirow{3}{*}{ Environmental } & $\mathrm{Cu}$ & $0.089(11.482 \%)$ & $0.095(9.546 \%)$ & $0.003(0.358 \%)$ & $-0.001(0.145 \%)$ \\
& $\mathrm{Pb}$ & $0.032(5.633 \%)$ & $0.038(5.109 \%)$ & $0.036(6.952 \%)$ & $0.004(5.803 \%)$ \\
& $\mathrm{Zn}$ & $0.097(18.306 \%)$ & $0.105(15.459 \%)$ & $-0.001(-0.194 \%)$ & $-0.002(-0.320 \%)$ \\
\multirow{2}{*}{ Socio- } & Melbourne & $0.150(50.444 \%)$ & $0.197(52.847 \%)$ & $-0.006(-2.079 \%)$ & $-0.008(-2.095 \%)$ \\
\multirow{2}{*}{ economic } & Sydney & $0.159(42.574 \%)$ & $0.229(47.299 \%)$ & $0.008(2.038 \%)$ & $0.034(7.548 \%)$ \\
& Brisbane & $0.087(20.176 \%)$ & $0.067(11.715 \%)$ & $0.019(5.087 \%)$ & $0.013(2.500 \%)$ \\
\hline
\end{tabular}
\label{table1}
\end{table}

\subsubsection{Performance in Non-geospatial models }

\begin{figure}[H]
\centering\includegraphics[width=1.0\linewidth]{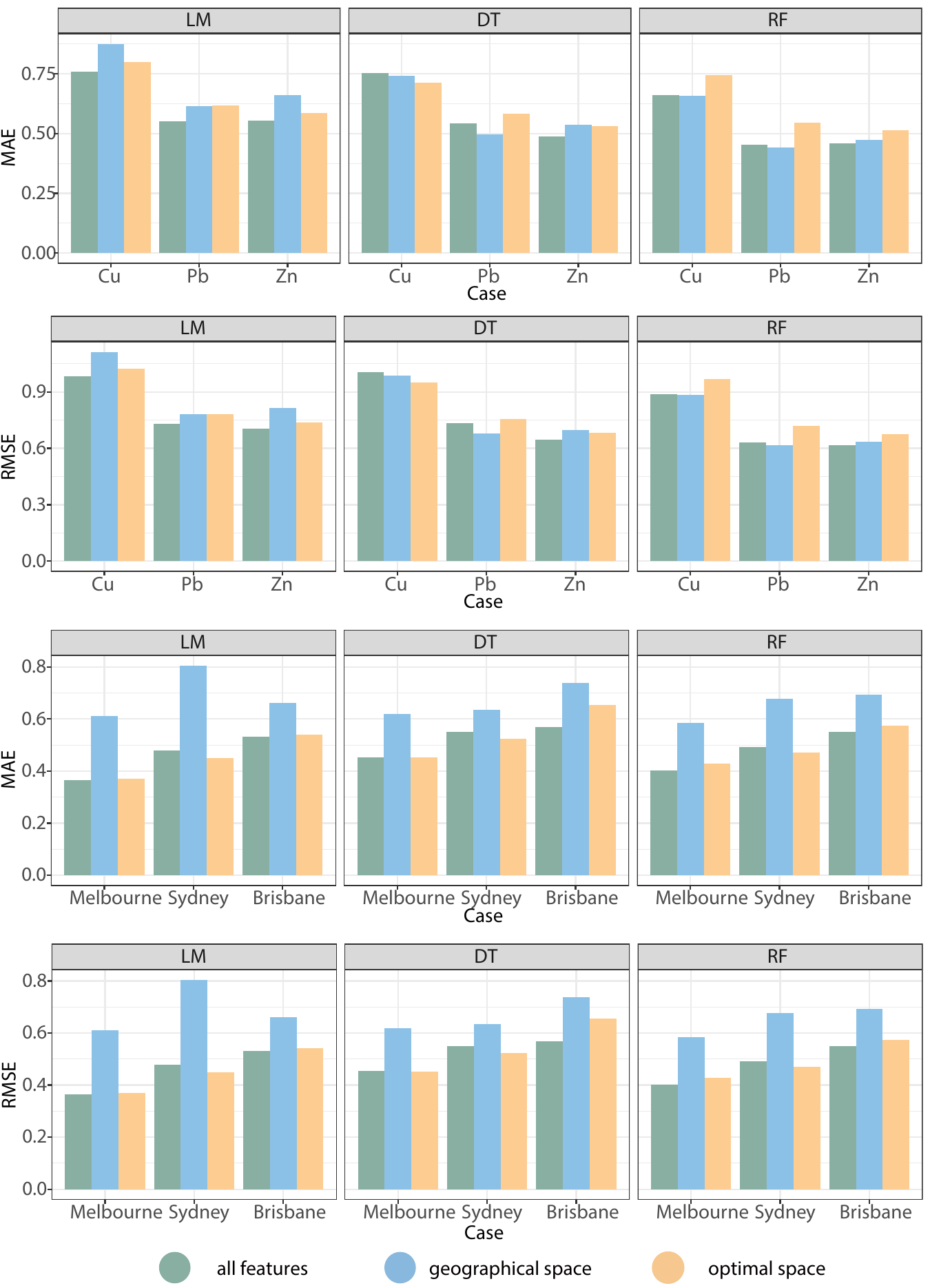} 
\caption{For non-geospatial models, predictive performance is compared between using all features (including geospatial features and attribute features), using only the two features included in geographical space (i.e., longitude and latitude), and using only the features included in the optimal space (the space with the strongest spatial dependence).}
\label{fig:07}
\end{figure}

Figure \ref{fig:07} illustrates the spatial prediction results of non-geospatial models under three distinct scenarios: employing all variables, using only geospatial variables, and utilizing the two attributes with the strongest spatial dependence. For environment variables, the LM and DT models demonstrate superior performance in the optimal attribute space over geographical variables, with the exception of one case involving Pb. However, for the RF model, the performance is poorest when utilizing the best spatial attributes. Regarding socioeconomic variables, all three models significantly outperform their geographical counterparts in the optimal attribute space. In many cases, the accuracy achieved by merely employing the two best spatial features even surpasses that of models utilizing all attributes.

In Supplementary Figure S3, a scatter plot of the true values versus the predicted values for each observation in leave-one-out cross-validation is displayed. The OK method demonstrates the poorest performance in geospatial contexts, with $R^2$ values ranging from 0.03 to 0.26. Its predictive range is confined to a narrow interval, not approximating the actual spatial distribution. This indicates that the spatial data in these six cases are insufficient to support the construction of a reasonable semivariogram. However, by transforming the geographic space into an optimal attribute space, there is a significant improvement in predictive performance. For environment variables, the enhancements in predictive performance for the three heavy metals Cu, Pb, Zn are 0.17, 0.12, 0.24, respectively. Regarding economic variables, improvements on the homeless rate in Melbourne, Sydney, Brisbane are 0.30, 0.36, 0.09, respectively. RK, as a method that employs all explanatory variables for trend surface fitting, shows a significant improvement over OK in geographic space. Yet, in most cases, its performance is comparable to that of OK using only the two optimal spatial dimensions. This underscores the importance of spatial dependence for the predictive performance of geospatial models. Moreover, in the optimal space, RK's predictions for the homeless rate in Sydney and Brisbane even outperform RF. In summary, the scatter plot results demonstrate that transforming geographic space into a attribute space with stronger spatial dependence allows geospatial models to more accurately capture the characteristics of spatial data, thereby making the model's output range closer to the true data distribution and improving predictive performance.

\section{Discussion and Conclusion}

Spatial data representation is crucial for understanding the geographical world and is central to predicting the distribution of geographical phenomena. Effective spatial data representation allows us to capture the essential properties and interrelations of geographical features, providing deep insights into the phenomena and processes within the complex world. In spatial data analysis, how representation techniques capture spatial dependence and heterogeneity is a vital area of research. Our study focuses on discussing spatial dependence.

Our research uncovered spatial dependence within non-geographic, attribute-based dimensions. Through comparative analysis of spatial and non-geospatial models, our findings reveal the widespread impact of spatial dependence on prediction outcomes across various domains. We demonstrated that spatial dependence significantly affects the prediction performance across all spaces, with the strongest spatial dependence typically found in geographic space for environmental variables, a trend that does not uniformly apply to socio-economic variables.

In spatial prediction tasks, mapping data to a feature space with strong spatial autocorrelation can facilitate the simplification of decision boundaries in prediction models, thereby enhancing prediction and classification efficiency. As illustrated in the upper part of Supplementary Figure S4, a positively autocorrelated feature space simplifies decision boundaries, which enhances the accuracy and stability of models on such data. Conversely, as demonstrated in the lower part of Supplementary Figure S4, a negatively autocorrelated feature space with complex and variable decision boundaries may increase the difficulty for models to process such data. The strength and direction (positive or negative) of spatial autocorrelation significantly influence the relationships among data points, thus determining the complexity of model learning.

In practical applications, a deep understanding and effective utilization of these spatial characteristics are crucial for designing prediction models that are tailored to specific data features. Our research reveals that the impact of geographic and non-geographic spaces on the performance of spatial predictions is universal. However, this does not imply that the characteristics of geographic data can be overlooked. Indeed, our analysis (see Figure 1) confirms that environmental variables, in particular, exhibit the strongest spatial dependencies in geographic spaces, indicating that geographic features can significantly assist prediction models in delineating clear decision boundaries.

Therefore, our study not only enriches the theoretical understanding of spatial data analysis but also provides new frameworks and tools for handling and analyzing complex spatial data in practice. By revealing the ubiquity and influence of spatial dependence, our research offers guidance for improving the accuracy and reliability of spatial predictions and opens new pathways for interdisciplinary research, aiding in better addressing the complex spatial issues of today's world.

Our results serve as an example of the contribution of the geography community to AI method, demonstrating how geographical knowledge and methods can provide a foundation for developing highly intelligent and adaptive systems. By deeply understanding the characteristics of spatial data, especially spatial dependence, we can not only improve the accuracy of spatial predictions but also lay the groundwork for building AI systems capable of understanding complex spatial relationships and patterns. This interdisciplinary integration aids the development of AI, as it requires learning from and interpreting complex phenomena in geographic spaces.

Additionally, the methods and findings of our study are significant for advancing technical innovation within the spatial data science and geographic information science fields. By integrating geographical spatial analysis with advanced machine learning techniques, we can develop more precise and flexible tools applicable to traditional geographical spatial predictions and wider applications, such as social network analysis, ecological modeling, and economic trend forecasting.

In summary, our research emphasizes the importance of considering spatial dependence in spatial data analysis, crucial for ensuring the accuracy and reliability of prediction models. This work not only promotes the development of spatial data processing techniques but also provides new perspectives for understanding and addressing complex systems, thereby advancing the field of artificial intelligence.

\section{Materials and Methods}

\subsection{Research data}

In this study, we have meticulously gathered and analyzed six key datasets to investigate the impact of spatial dependence on spatial predictions (Supplementary Figure S1). Specifically, our datasets encompass detailed distributions of trace elements within a region in Australia, as well as the distribution of homelessness rates across three major cities—Brisbane, Melbourne, and Sydney. 

For the distribution of trace elements, we focused on the spatial distribution of  Cu, Zn, and Pb. These elements are widely considered significant indicators of environmental pollution and are crucial for assessing environmental health. To explain the spatial variability of these trace elements, we collected nine environment-related explanatory variables, including distance to lithology (Dlith), distance to fault (Dfault), slope, water distribution, Normalized Difference Vegetation Index (NDVI), distance to main roads (MainRd), road network density (Road), soil organic carbon (SOC), and soil pH value. These explanatory variables are deemed essential for understanding and predicting the distribution of trace elements. Detailed distributions of these data can be found in the Supplementary Figure S1.

Regarding socio-economic indicators, we identified the homelessness rate as a key measure of socio-economic status. The homelessness rate not only reflects the economic challenges faced by individuals and families but also indicates the effectiveness of social governance and support systems. We collected homelessness rate data at the SA3 regional level for Australia's three major cities—Brisbane, Melbourne, and Sydney. To thoroughly analyze the factors affecting the homelessness rate, we likewise selected nine explanatory variables, including population density (popdens), average income (avgincome), health status (health), unemployment rate (unemploy), lack of internet access (nointernet), rental payment capacity (rentalpay), mortgage pressure (mortage), proportion of rented housing (rented), affordability of mortgages (affordmort), and commute distance to work (distwork). These variables are considered valuable in assessing and understanding socio-economic conditions and their impact on homelessness rates. 

Supplementary Figure S2 displays the statistical distribution and Q-Q plots of six sets of data after logarithmic transformation. These six sets of data exhibit skewness issues in their original states, indicating deviations from a normal distribution. However, once subjected to logarithmic transformation, the distributions of these data tend to become more normal. It is important to note that, compared to the trace element data, the data on the risk of homelessness in families seem to have more outliers. Particularly in Brisbane, we can clearly observe the presence of an outlier, which may be attributed to the socioeconomic attribute data in that region being influenced by more anthropogenic factors.

\subsection{Experiment design}

\begin{figure}[H]
\centering\includegraphics[width=1.0\linewidth]{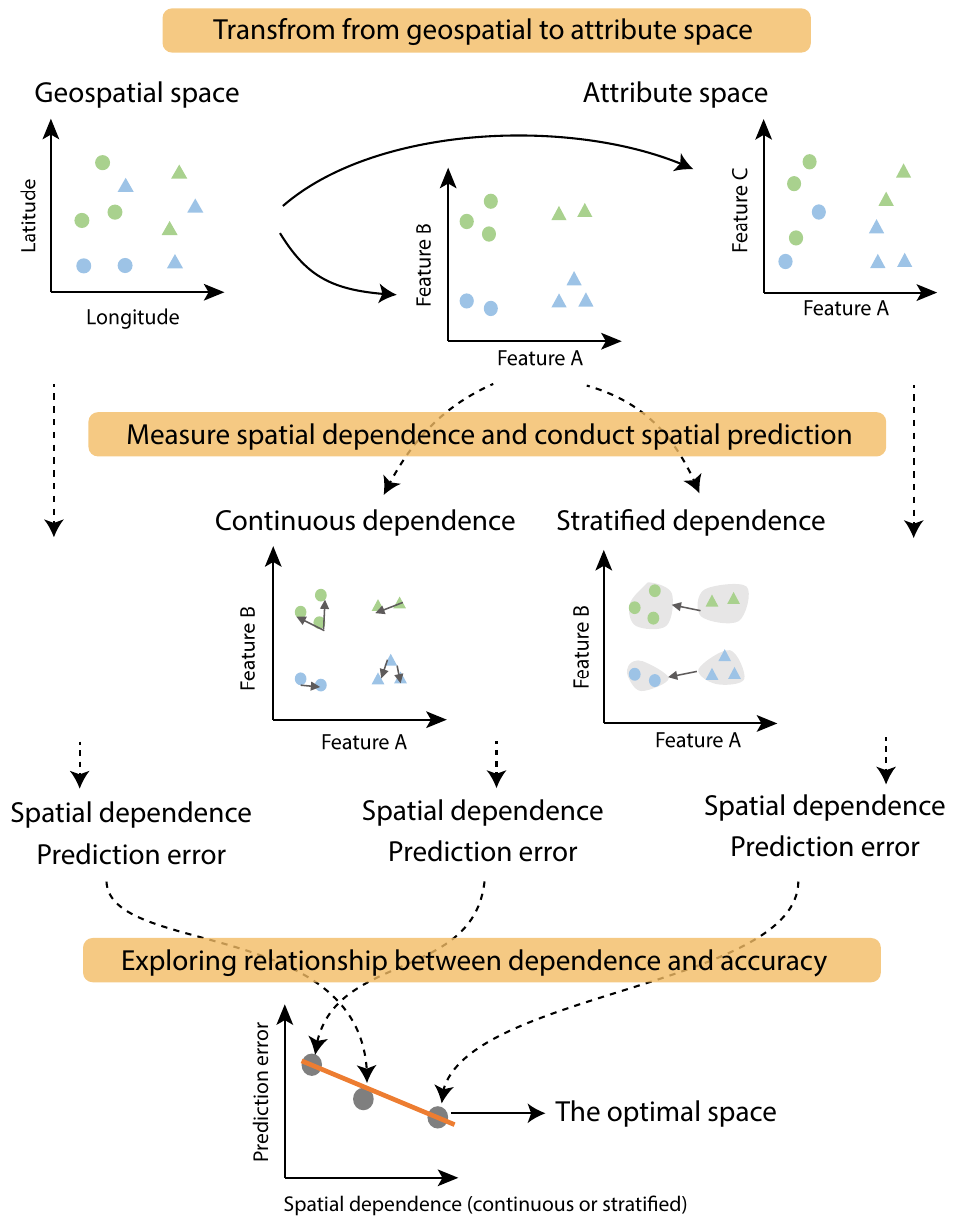} 
\caption{Experiment design to test the relationship between prediction performance and spatial dependence}
\label{fig:08}
\end{figure}

This study treats spatial prediction as a general term that encompasses both geographic interpolation and extrapolation. We investigate how generalized spatial prediction methods perform in relation to both environmental and socio-economic variables. To achieve this, we selected trace elements as an example of environmental variables and homelessness as an example of socio-economic variables. We intentionally chose not to differentiate between datasets based on whether they are discrete or continuous, focusing instead on their spatial characteristics, allowing us to explore the broader applicability of spatial prediction models across different fields and data types.

Figure \ref{fig:08}  shows our experimental design, which aims to reveal the relationship between spatial dependence and the performance of spatial prediction. We collected data for a geographic variable X from geospatial areas, arranged in space by latitude and longitude. Assuming that for the collected data locations, we can obtain the values of three other geographic variables, namely A, B, and C. In the example shown in the figure, the attributes can be color and shape. We first transform the collected geospatial data of the geographic variable from its spatial arrangement to a space composed of any two attributes. Then, in the attribute space, we calculate the spatial dependence of geographic variable X. In this study, we consider two types of spatial dependence: stratified dependence and continuous dependence. Concurrently, we perform spatial prediction for X in this space and calculate its accuracy. Thus, for any combination of two attributes, including combinations with longitude or latitude, we calculate their prediction errors and spatial dependencies. In the final step, based on the predictive accuracy and dependencies obtained for all spaces, we can explore the relationship between the two. In this study, we use the Pearson correlation coefficient for analysis.

It should be mentioned that Moran's I index are normally applied to discrete spatial data. For both trace elements and homelessness data, we consider the datasets to be spatially discrete. As such, using the Moran’s I index to assess their spatial autocorrelation is appropriate. Specifically, for predicting trace elements, applying Moran’s I to analyze their spatial patterns is a common and valuable practice. For instance, LISA can help identify hotspots in the spatial distribution of a specific metal concentration. While geostatistics models are frequently applied to trace elements, it is uncommon to see them used for predicting socio-economic attributes such as homelessness. However, in many real-world applications, census-derived statistical data is often used to fill in missing values, especially when there are insufficient explanatory variables for a regression-based prediction model. In these cases, interpolation techniques, including geostatistics, are often employed.

\subsection{Indicators of spatial dependence}

The continuous dependence is measured by the Moran's I inedx:

\begin{equation}
I = \frac{N}{\sum_{i=1}^{N}\sum_{j=1}^{N}w_{ij}} \frac{\sum_{i=1}^{N}\sum_{j=1}^{N}w_{ij}(X_i - \bar{X})(X_j - \bar{X})}{\sum_{i=1}^{N}(X_i - \bar{X})^2}
\end{equation}
where $I$ represents the value of Moran's I index, $N$ is the total number of observations, $X_i$ and $X_j$ denote the attribute values of observations $i$ and $j$, respectively. $\bar{X}$ is the mean of all observation attribute values. $w_{ij}$ is the spatial weight between observations $i$ and $j$ in the spatial weight matrix. The denominator, $\sum_{i=1}^{N}(X_i - \bar{X})^2$, is the sum of squared deviations of the attribute values. The numerator, $\sum_{i=1}^{N}\sum_{j=1}^{N}w_{ij}(X_i - \bar{X})(X_j - \bar{X})$, is the sum of the product of the weighted deviations of the attribute values.

The stratified dependence is measured by the q value:

\begin{equation}
q=\min \left(S S W_{X, D}\right)=\min \left\{\sum_{z=1}^h \sum_{j=1}^{N_z}\left(y_{z, j}-\bar{c}_z\right)^2\right\}
\end{equation}
where $X$ represents one or several predictors, $D$ delineates the variable for stratification based on geographical segments, and $SSW_{X, D}$ denotes the within-stratum sum of squares, attributed to the geographical segments identified by $D$ and influenced by the predictor(s) $X$. The notation $y_{z,j}$ signifies the $j$th measurement of the dependent variables within the $z$th stratum, while $\overline{c}_{z}$ refers to the average values of these dependent variables in each respective stratum $z$.



\subsection{Prediction model}

Five models were selected for spatial prediction of trace elements and homeless rates, including two spatial interpolation models, Ordinary Kriging (OK) and Regression Kriging (RK), along with three non-geospatial models, Linear Regression (LR), Decision Trees (DT), and Random Forests (RF). To explore the relationship between the performance of spatial predictions and spatial dependence, we randomly divided each dataset into an 80\% training set and a 20\% test set. Models were trained on the training set and then validated on the test set. Root Mean Square Error (RMSE) and Mean Absolute Error (MAE) were chosen as accuracy assessment metrics. For each experimental setup, we repeated the process 50 times and calculated the average of the accuracy metrics as the final accuracy result. In analyzing the quantitative impact of spatial dependence on the performance of prediction models, we employed leave-one-out cross-validation. Specifically, each point in the dataset was used as the test set in turn, with the remaining data serving as the training set. Accuracy was then assessed using RMSE and MAE.

\section*{Data available statement}

The data and codes that support the findings of the present study are available on Figshare.

\section*{Disclosure Statement}

No conflict of interest exists in this manuscript, and the manuscript was approved by all authors for publication.

\baselineskip12pt
\bibliographystyle{elsarticle-harv} 
\bibliography{02_references.bib}


\end{document}


\begin{frontmatter}
\title{Supplementary}


\end{frontmatter}

\linenumbers




\begin{figure}[ht!]
\centering\includegraphics[width=1.0\linewidth]{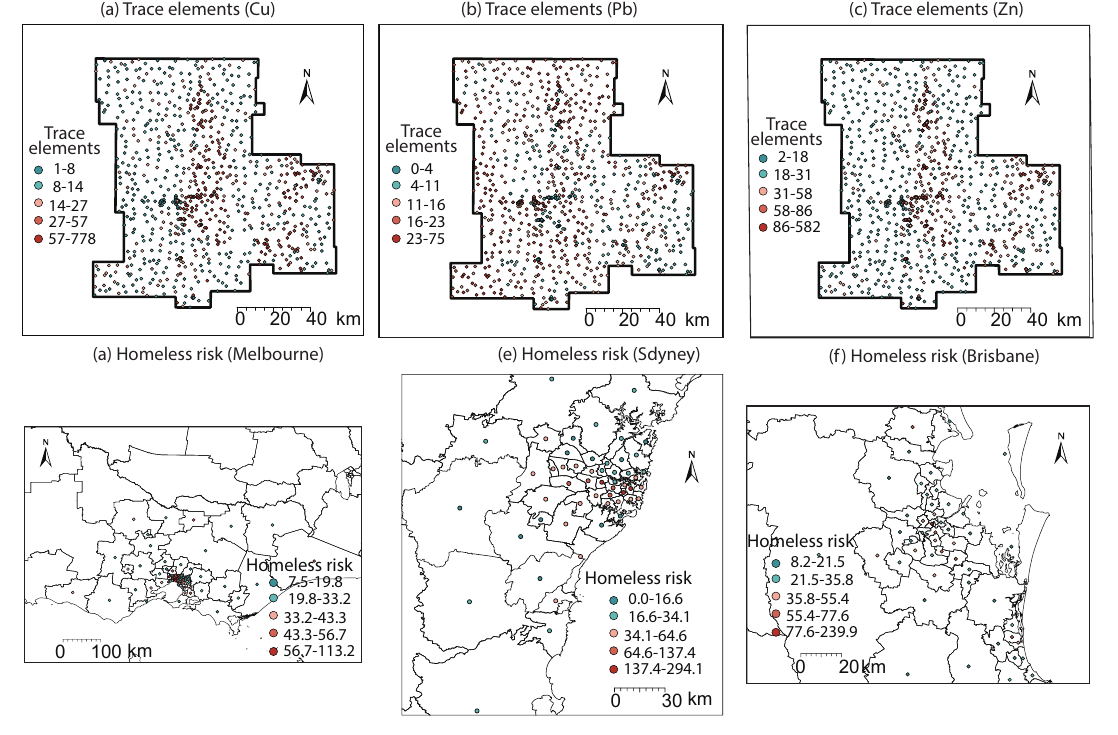}
\caption{Datasets: (a-c) Trace elements, (d-f) Homeless risk}
\label{fig:a1}
\end{figure}

\begin{figure}[ht!]
\centering\includegraphics[width=1.0\linewidth]{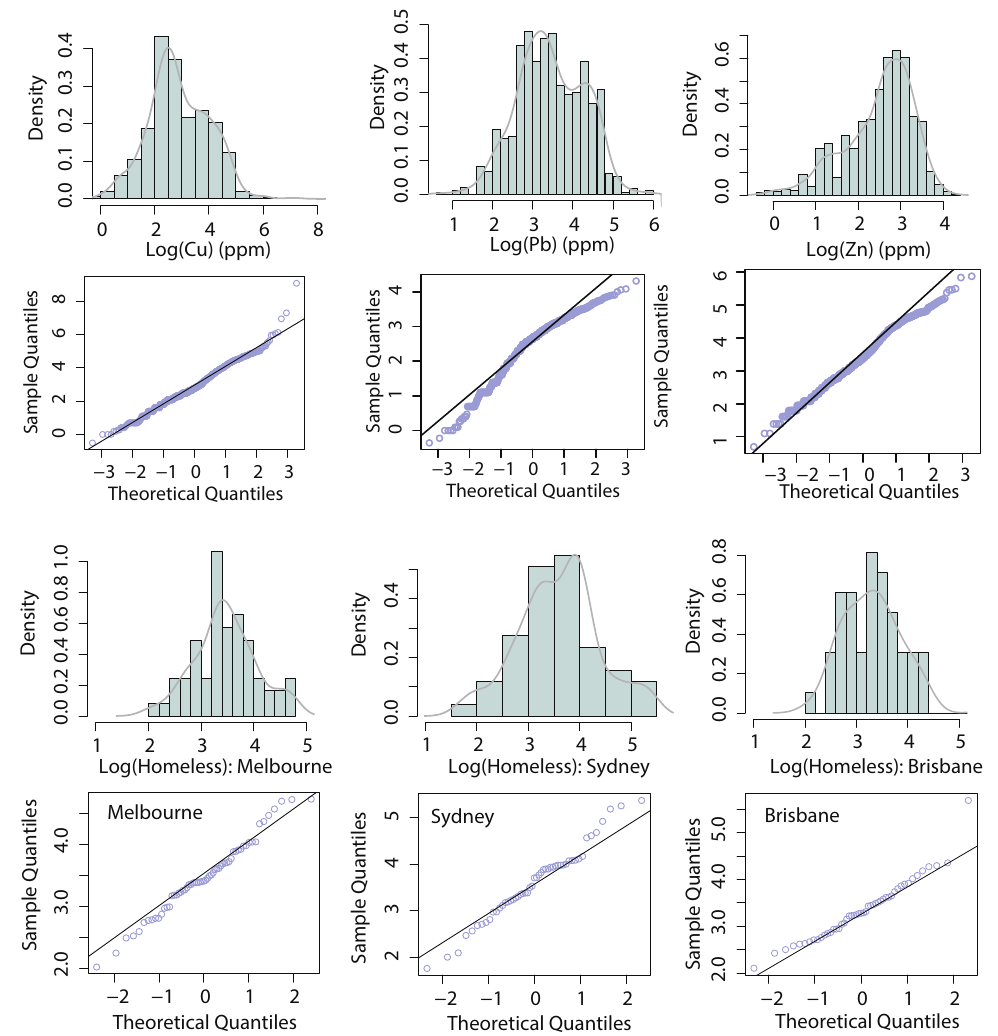}
\caption{Data distribution after pre-processing}
\label{fig:a2}
\end{figure}



\begin{table}[ht!]
\centering
\scriptsize
\caption{Explanatory variables of the homelessness risk}
\begin{tabular}{ccccccc}
\hline
Name & Code & Description \\
\hline
average income & avgincome & Mean employee income (\$)\\
property mortgage & mortage & Owned property with a mortgage (\%)\\
rented property & rented & Rented property percentage (\%) \\
health people & healthpop & Persons with long-term health conditions (\%) \\
population density & popdens & population density (persons/$km^2$) \\
unemployment rate & unemployment & unemployment rate (\%) \\
non internet rate & noninternet & proportion of dwelling without Internet access (\%) \\
rental payment & rentalpay & median weekly household rental payment $(\$)$ \\
mortgage affordability & affordmort & $\begin{array}{c}\text { Households where mortgage repayments are } \\
\text { more than  30\% of imputed household income }\end{array}$ \\
distance to work & diswork & $\begin{array}{c}\text {average commuting distance from place of usual } \\
\text {residence to work (km) }\end{array}$ \\
\hline
\end{tabular}
\label{table:kriging}
\end{table}


\begin{table}[ht!]
\centering
\scriptsize
\caption{Explanatory variables of the trace elements}
\begin{tabular}{ccccccc}
\hline
 Code & Description \\
\hline
  Dlith & Distance to trace element-related lithology (km)\\
  Dfault &  Distance to trace element-related fault  (km)\\
  Slope & Slope (\%) \\
  Water & Distance to water, including rivers and lakes (km) \\
  NDVI & Normalized difference vegetation index \\
  MainRd & unemployment rate (\%) \\
  Road & distance to roads, including state and local roads (km) \\
  SOC & soil organic carbon \\
  pH & pH \\
\hline
\end{tabular}
\label{table:kriging}
\end{table}


\begin{figure}[ht!]
\centering\includegraphics[width=1.0\linewidth]{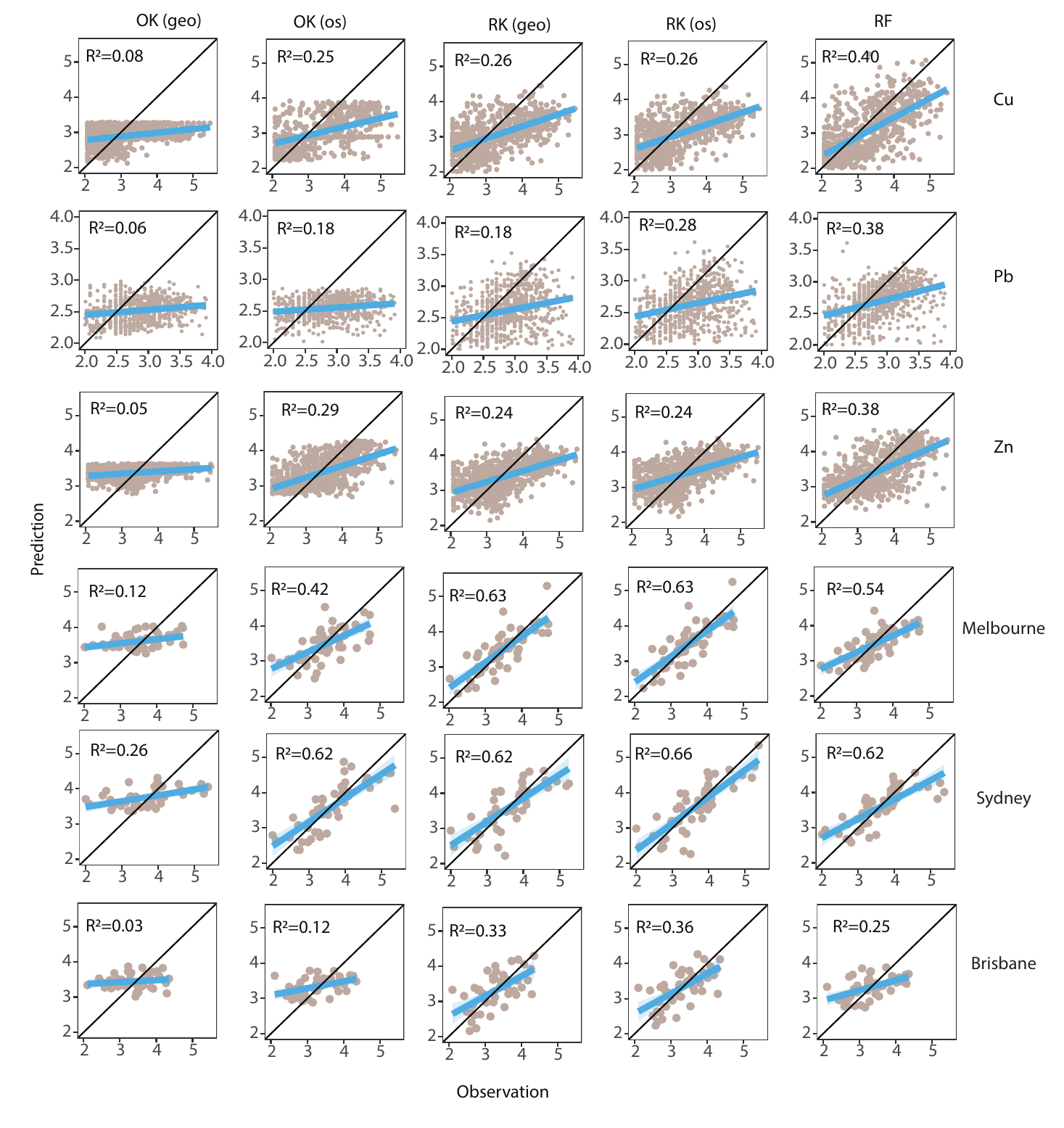}
\caption{Comparisons between observations and predictions for five models}
\label{fig:a3}
\end{figure}

\begin{table}[!ht]
    \centering
    \caption{Performance of non-spatial models at different space (Environment variables)}
\begin{tabular}{cc|rc|cc|cc}
\hline &  & \multicolumn{2}{|c|}{ Cu } & \multicolumn{2}{c|}{ Pb } & \multicolumn{2}{c}{ Zn } \\
\hline Methods & Features & \multicolumn{1}{|c|}{ RMSE } & MAE & RMSE & MAE & RMSE & MAE \\
\hline \multirow{4}{*}{ LM } & all & 0.985 & 0.760 & 0.729 & 0.552 & 0.703 & 0.555 \\
 & geo & 1.112 & 0.876 & 0.780 & 0.614 & 0.814 & 0.660 \\
 & os & 1.023 & 0.799 & 0.780 & 0.618 & 0.738 & 0.585 \\
 & all & 1.004 & 0.752 & 0.733 & 0.543 & 0.646 & 0.489 \\
DT & geo & 0.985 & 0.741 & 0.677 & 0.496 & 0.697 & 0.537 \\
 & os & 0.949 & 0.714 & 0.756 & 0.583 & 0.682 & 0.531 \\
 & all & 0.887 & 0.661 & 0.630 & 0.453 & 0.618 & 0.460 \\
RF & geo & 0.886 & 0.658 & 0.615 & 0.443 & 0.634 & 0.473 \\
 & os & 0.969 & 0.744 & 0.720 & 0.546 & 0.675 & 0.513 \\
\hline
\end{tabular}
\label{table2:}
\end{table}

\begin{table}[!ht]
    \centering
    \caption{Performance of non-spatial models at different space (Socio-economic variables)}
\begin{tabular}{cc|cc|cc|cc}
\hline & & \multicolumn{2}{|c|}{ Melbourne } & \multicolumn{2}{c|}{ Sydney } & \multicolumn{2}{c}{ Brisbane } \\
\hline Methods & Features & RMSE & MAE & RMSE & MAE & RMSE & MAE \\
\hline \multirow{4}{*}{ LM } & all & 0.365 & 0.263 & 0.478 & 0.369 & 0.532 & 0.394 \\
& geo & 0.611 & 0.478 & 0.804 & 0.645 & 0.662 & 0.514 \\
& os & 0.370 & 0.296 & 0.449 & 0.348 & 0.541 & 0.408 \\
& all & 0.454 & 0.366 & 0.551 & 0.465 & 0.569 & 0.443 \\
DT & geo & 0.619 & 0.486 & 0.635 & 0.532 & 0.739 & 0.574 \\
& os & 0.452 & 0.384 & 0.522 & 0.441 & 0.655 & 0.472 \\
& all & 0.401 & 0.322 & 0.492 & 0.372 & 0.550 & 0.428 \\
RF & geo & 0.585 & 0.464 & 0.677 & 0.539 & 0.694 & 0.535 \\
& os & 0.429 & 0.353 & 0.471 & 0.360 & 0.574 & 0.418 \\
\hline
\end{tabular}
\label{table3}
\end{table}


\begin{figure}[ht!]
\centering\includegraphics[width=1.0\linewidth]{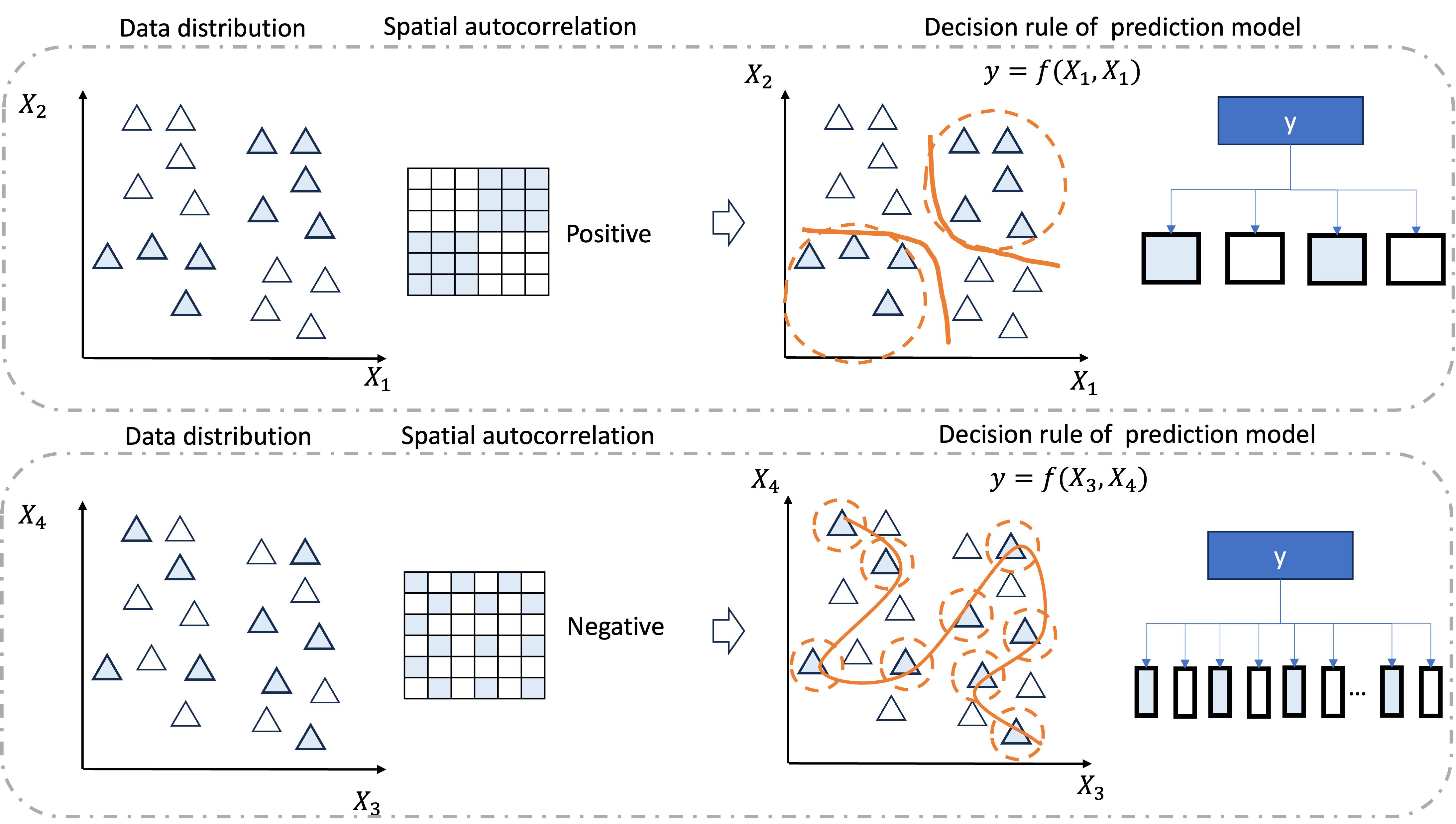}
\caption{Positive spatial dependence facilitates simpler decision rules in prediction models}
\label{fig:a2}
\end{figure}